\documentstyle[preprint,prd,aps,epsf]{revtex}
\begin{document}
\draft

\title{Nonlinear theory of flame front instability}

\author{Kirill~A.~Kazakov${ \mbox{}}^{1,2}$\thanks{E-mail: $Kirill.Kazakov@fysik.uu.se$} and
Michael~A.~Liberman${ \mbox{}}^{1,3}$\thanks{E-mail:
$Michael.Liberman@fysik.uu.se$} }

\address{${\mbox{}}^{1}$Department of Physics, Uppsala University, Box
530, S-751 21, Uppsala, Sweden \\
${\mbox{}}^{2}$ Moscow State University, Physics Faculty, Department of Theoretical Physics, \\
117234, Moscow, Russian Federation
\\ ${\mbox{}}^{3}$P.~Kapitsa Institute
for Physical  Problems, Russian Academy of Sciences,  \\
117334, Moscow, Russian Federation}

\maketitle

\tightenlines

\begin{abstract}
Nonlinear non-stationary equation describing evolution of weakly
curved premixed flames with arbitrary gas expansion, subject to
the Landau-Darrieus instability, is derived. The new equation
respects all the conservation laws to be satisfied across the
flame front, as well as correctly takes into account influence of
vorticity, generated in the flame, on the flame front structure
and flame velocity. Analytical solutions of the derived equation
are found.
\end{abstract}


\narrowtext

\unitlength=1pt

\noindent

\section{Introduction}

Description of premixed flame propagation is, in essential, the description
of development of the Landau-Darrieus (LD) instability \cite{landau,darrieus}
of zero-thickness flames. Given arbitrary flame front configuration,
its evolution is determined by the exponential growth of unstable modes,
eventually stabilized by the nonlinear mode interaction. By themselves,
the nonlinear effects are not sufficient to stabilize flame propagation,
since the spectrum of unstable perturbations of a zero-thickness flame
is unbounded. In many cases, however, an upper bound for the mode wavenumber
is provided by the heat conduction -- species diffusion processes in the
flame, which govern the evolution of short-wavelength perturbations
\cite{pelce}. The flame propagation is thus described as the nonlinear
propagation of interacting modes of zero-thickness front, with an effective
short-wavelength cut-off described at the hydrodynamic scale by means of an
appropriate modification of the evolution equation and the conservation
laws at the flame front \cite{matalon}.

In this purely hydrodynamic formulation, the flame dynamics is
governed essentially by the only parameter -- the gas expansion
coefficient $\theta,$ defined as the ratio of the fuel density and
the density of burnt matter. Unfortunately, in general it is very
difficult to reduce the complete system of hydrodynamic equations
governing flame dynamics to a single equation for the flame front
position. This is mainly because the nonlinearity of flame
dynamics cannot be considered perturbatively. For instance, it can
be shown that in the regime of steady flame propagation, the flame
front slope can be considered small only if the gas expansion is
small ($\theta \to 1$), while for flames with $\theta = 6\div 8$
it is of the order $2\div 3$ (discussion of this issue can be
found in Ref.~\cite{kazakov}). At the early stages of development
of the LD-instability, however, the perturbation analysis is fully
justified, and the equation describing nonlinear propagation of
the flame front can be obtained in a closed form. Within accuracy
of the second order in the flame front slope such an equation was
obtained by Zhdanov and Trubnikov (ZT) \cite{zhdanov}, without
taking into account the influence of the effects related to finite
flame thickness, mentioned above. The latter were included in the
ZT-equation {\it ad hoc} by Joulin \cite{joulin}.

Concerning the ZT-equation and its modification, we would like to note
the following.

1) Although this equation respects all the conservation laws to be
satisfied across the flame front, it was derived on the basis of a
certain model assumption concerning the flow structure downstream.
Namely, it was assumed that the velocity field can be represented
as a superposition of a potential mode and an ``entropy wave'', so
that the pressure field is expressed through the former by the
usual Bernoulli equation. The generally nonlocal relation between
the pressure and velocity fields is thus rendered {\it local
algebraic}, which allows simple reduction of the system of
hydrodynamic equations to the single equation for the flame front
position. Being valid at the linear stage of development of the
LD-instability, this model assumption is, of course, unjustified
in general.

2) It was assumed in the course of derivation of the ZT-equation
that not only the front slope, but also the value of the front
position itself is a small quantity. As a result, the ZT-equation
turns out to be non-invariant with respect to space translations
in the direction of flame propagation. Practical consequence of
this assumption is the unnecessary limitation of the range of
validity of the equation: the space and time intervals should be
taken sufficiently small to ensure that the deviation of the front
position from the initial unperturbed plane is small everywhere.

The purpose of this paper is to show that the only assumption of
smallness of the flame front slope is actually sufficient to
derive an equation describing the nonlinear development of the
LD-instability to the leading (second) order of nonlinearity.
Surprisingly, this equation turns out to be of a more simple
structure than that of the ZT-equation. This simplification is due
to existence of a representation of the flow equations at the
second order, which can be called {\it transverse}. In this
representation, the system of equations governing the flame
propagation can be brought into the form in which dependence of
all dynamical quantities on the coordinate in the direction of
flame propagation is rendered purely parametric.

Let us consider the question of existence of this representation,
and more generally, the meaning of the weak nonlinearity
expansion, in more detail. First of all, the following important
aspect of the problem should be emphasized. The curved flame
propagation is an essentially nonlocal process, in that the
presence of vorticity produced in the flame implies that the
relations between flow variables downstream generally cannot be
put into the form in which the value of one variable at the front
surface can be expressed entirely in terms of other variables
taken at the same surface. For instance, the value of the pressure
field at the flame front depends not only on the gas velocity
distribution along the front, but also on its distribution in the
bulk. This means in turn that the equation describing the flame
front evolution cannot generally be written in a closed form, {\it
i.e.,} as an equation which expresses the time derivative of the
front position via its spatial gradients, because the gas dynamics
in the bulk depends, in particular, on the boundary conditions for
the burnt matter. In the framework of the weak nonlinearity
expansion, this non-locality shows itself as the necessity to
increase the differential order of the equation for the flame
front position. For example, it was shown in Ref.~\cite{kazakov}
that in order to take into account influence of the vorticity
drift on the structure of stationary flames with the accuracy of
$(\theta - 1)^4$ in the asymptotic expansion for $\theta \to 1$
(which is the relevant weak nonlinearity expansion in the
stationary case), one has to increase the differential order of
the integro-differential equation by one as compared to the
Sivashinsky equation (the latter is of the second order in $\theta
- 1,$ corresponding to the approximation in which the gas flow is
potential on both sides of the front). Roughly speaking, the
nonlocal relations, being {\it integral} from the non-perturbative
point of view, are treated {\it differential} of infinite order in
the framework of the weak nonlinearity expansion.

Let us now turn back to our present purpose of derivation of
equation for the front position at the second order of
nonlinearity. Remarkably, it turns out that this approximation is
exceptional in that the above-mentioned nonlocal complications do
not arise in this case. We will give now a simple illustration of
this important fact. Let us consider a weakly curved flame
propagating in $z$-direction with respect to an initially uniform
fuel, and denote ${\bf x}$ the transverse coordinates. If the
flame front position is described by equation $z = f({\bf x},t),$
then the condition that front is only weakly curved implies that
$\partial f({\bf x},t)/\partial{\bf x}$ is small, and so is the
gas velocity perturbation $\delta {\bf v}$. From the continuity
equation $${\rm div}{\bf v} = 0$$ and the Euler equations
$$\frac{\partial {\bf v}}{\partial t} + ({\bf v}\nabla){\bf v}
= - \frac{1}{\rho}\nabla P,$$ one has
\begin{eqnarray}\label{intr1}
\nabla^2 P = - \rho~
\partial_i v_k \partial_k v_i\,,
\end{eqnarray} \noindent where summation over repeated indices is
understood. Equation (\ref{intr1}) implies
\begin{eqnarray}\label{intr2}
P(z,{\bf x},t) = - \rho~\int dz' d{\bf x}' G(z,{\bf x},z',{\bf
x}',t)
\partial_i v_k (z',{\bf x}',t) \partial_k v_i (z',{\bf x}',t)
+ \Omega (z,{\bf x},t) \,,
\end{eqnarray} \noindent where $\Omega (z,{\bf x},t)$ is a local
function of the flow variables, satisfying $\nabla^2 \Omega = 0,$
and $G(z,{\bf x},z',{\bf x}',t)$ is the Green function of the
Laplace operator, appropriate to the given boundary condition.
Note that the latter, being a condition on the pressure jump
across the flame, is imposed only at the flame front. Since the
unperturbed velocity field is spatially uniform, one has from
Eq.~(\ref{intr2}) for the curvature induced pressure variation
$\delta P$
\begin{eqnarray}\label{intr3}
\delta P(z,{\bf x},t) = - \rho~\int dz' d{\bf x}' G(z,{\bf
x},z',{\bf x}',t)
\partial_i \delta v_k (z',{\bf x}',t) \partial_k \delta v_i (z',{\bf
x}',t) + \delta\Omega (z,{\bf x},t)\,.
\end{eqnarray}
\noindent Next, let us denote $\delta_x$ the characteristic length
of the front perturbation. Then the corresponding length in
$z$-direction $$\delta_z \sim \left|\frac{\partial f}{\partial{\bf
x }}\right|\delta_x\,.$$ In other words, integration over $z'$ in
Eq.~(\ref{intr3}) is effectively carried out for $z'\in (f({\bf
x},t), f({\bf x},t) + \delta_z).$ Furthermore, as we will see in
the sections below, $$|\delta{\bf v}| \sim \left|\frac{\partial
f}{\partial t}\right| \sim \left|\frac{\partial f}{\partial{\bf x
}}\right|.$$ Therefore, if one is interested in evaluating the
pressure variation at the front, one can rewrite Eq.~(\ref{intr3})
with the accuracy of the second order
\begin{eqnarray}\label{intr4}
\delta P(f({\bf x},t),{\bf x},t) &=& - \rho~\int d{\bf x}'
\tilde{G}({\bf x},{\bf x}',t)
\partial_i \delta v_k (f({\bf x}',t),{\bf x}',t) \partial_k
\delta v_i (f({\bf x}',t),{\bf x}',t) \nonumber\\ &+& \delta\Omega
(f({\bf x},t),{\bf x},t)\,,\\ ~~ \tilde{G}({\bf x},{\bf x}',t)
&\equiv& \int dz' G(f({\bf x},t),{\bf x},z',{\bf
x}',t)\,.\nonumber
\end{eqnarray}
\noindent As we noted above, ``boundary condition'' for the
pressure field is imposed only at the flame front. Therefore, the
Green function $\tilde{G}$ is independent of any other conditions
relevant to the flow of the burnt matter ({\it e.g.,} boundary
conditions on the tube walls, in the case of flame propagation in
a tube). The same is true for the function $\Omega,$ since the
value of $\Omega$ at a given point depends only on the value of
flow variables at the same point. We thus see that
Eq.~(\ref{intr4}) is an integral relation between local functions
of the flow variables, defined on the front surface, which is
independent of the flow dynamics in the bulk. Furthermore, since
the right hand side of Eq.~(\ref{intr4}) is of the second order,
it is not difficult to show, using the linear decomposition of the
flow field into potential and vortex components, that
$z$-derivatives of the velocity can be expressed via its
derivatives along the flame front, bringing this equation into the
transverse representation. This implies that at the second order,
there exists a universal equation which describes the flame front
dynamics in terms of the front position alone. In practice, it is
actually more convenient to work with the differential form of the
flow equations, rather than integral. How their transverse
representation can be derived will be shown in detail in
Sec.~\ref{transvers}. On the basis of this result, the nonlinear
non-stationary equation will be derived in Sec.~\ref{nnequation}.
Its analysis is carried out analytically using the method of pole
decomposition in~Sec.~\ref{decomposition}. The obtained results
are summarized in Sec.~\ref{conclud}.

\section{Flow equations and conservation laws}\label{general}

Let us consider a 2D flame propagating in the negative
$z$-direction with the normal velocity $U_{\rm f}$ with respect to
an initially uniform quiescent combustible gas mixture. Denoting
$x$ the transverse coordinate, we introduce dimensionless space
and time variables $(\eta,\xi)\equiv (x/R,z/R),$ $\tau \equiv t
U_{\rm f}/R,$ where $R$ is a characteristic gasdynamic length of
the problem. Then the normalized fluid velocity ${\bf v} =
(v_x/U_{\rm f},v_z/U_{\rm f}) \equiv (w,u)$ and pressure $\Pi = (P
- P(\xi = - \infty))/\rho_{-} U_{\rm f}^2$ obey the following
equations in the bulk
\begin{eqnarray}
\frac{\partial u}{\partial\xi} + \frac{\partial w}{\partial\eta}
&=& 0 \,,\label{flow1}\\
\frac{\partial u}{\partial\tau} + u
\frac{\partial u}{\partial\xi} + w \frac{\partial u}{\partial\eta}
&=& - \frac{1}{\varrho}\frac{\partial
\Pi}{\partial\xi}\,,\label{flow2}\\
\frac{\partial w}{\partial\tau} + u \frac{\partial w}{\partial\xi}
+ w \frac{\partial w}{\partial\eta} &=& -
\frac{1}{\varrho}\frac{\partial \Pi}{\partial\eta}\,,\label{flow3}
\end{eqnarray}
\noindent
\noindent where $\varrho$ is the fluid density scaled on the
density $\rho_{-}$ of the fuel.

The above flow equations are complemented by the following
conservation laws to be satisfied across the flame front
\cite{matalon}
\begin{eqnarray}
u_{+} - u_{-} - \frac{\partial f}{\partial\eta}(w_{+} - w_{-}) &=&
(\theta - 1) N\ ,\label{conserv1}\\
w_{+} - w_{-} + \frac{\partial f}{\partial\eta}(u_{+} - u_{-}) &=&
\varepsilon\ln\theta\left(\hat{D}w_{-} + \frac{\partial f
}{\partial\eta}\hat{D}u_{-} + \frac{1}{N}\hat{D}\frac{\partial
f}{\partial\eta} \right)\,,\label{conserv2}\\
\Pi_{+} - \Pi_{-} &=& - (\theta - 1) + \varepsilon (\theta -
1)\frac{\partial}{\partial\eta}\left(\frac{1}{N}\frac{\partial
f}{\partial\eta}\right) \nonumber\\ +
\frac{\varepsilon\ln\theta}{N}\left(\frac{\partial^2 f}{\partial
\tau^2} + 2w_{-}\frac{\partial^2 f}{\partial\tau
\partial\eta} \right.&+&\left. w_{-}^2 \frac{\partial^2 f}{\partial\eta^2} + 2 \hat{D}N -
\frac{1}{N}\frac{\partial f}{\partial\eta}\frac{\partial N
}{\partial\eta}\right)\,,\label{conserv3}
\end{eqnarray}
\noindent where
$$\hat{D}\equiv \frac{\partial}{\partial\tau} +
\left(w_{-} + \frac{1}{N}\frac{\partial
f}{\partial\eta}\right)\frac{\partial}{\partial\eta}\,, ~~N \equiv
\sqrt{1 + \left(\frac{\partial f}{\partial\eta}\right)^2}\ ,$$ and
$\varepsilon$ is the small dimensionless ratio of the flame
thickness to the characteristic length.

Finally, the following so-called evolution equation
\begin{eqnarray}\label{evolution}&& u_{-} - \frac{\partial f}{\partial\eta}w_{-}
- \frac{\partial f}{\partial\tau}
= N - \varepsilon\frac{\theta\ln\theta}{\theta -
1}\left(\frac{\partial N}{\partial\tau} +
\frac{\partial}{\partial\eta}(N w_{-}) + \frac{\partial^2 f
}{\partial\eta^2}\right),
\end{eqnarray}
\noindent completes the above system of hydrodynamic equations and
conservation laws, determining dynamics of the flame front itself.

Below, we will need the general solution of the flow equations
(\ref{flow1})--(\ref{flow3}) upstream. Going over to the rest frame of
reference of the initially plane flame front, this solution is readily
obtained as follows. Since the flow is potential at $\xi = - \infty$
(where $u = 1,$ $w = 0$), it is potential for every $\xi<f(\eta,\tau)$
in view of the Thomson theorem \cite{landafshitz}, thus
\begin{eqnarray}&&\label{solup1}
u \equiv 1 + \tilde{u} = 1 + \int\limits_{- \infty}^{+ \infty}d k
~\tilde{u}_k \exp (|k|\xi + i k\eta)\,,
\end{eqnarray}
\begin{eqnarray}&&\label{solup2}
w = \hat{H}\tilde{u}\,,
\end{eqnarray}
\begin{eqnarray}&&\label{solup3}
\frac{\partial \tilde{u}}{\partial\tau} + \hat{\Phi}\Pi +
\frac{\hat{\Phi}}{2}(u^2 + w^2) = 0\,,
\end{eqnarray}
\noindent where the Hilbert operator $\hat{H}$ is defined by
$$(\hat{H}f)(\eta) = \frac{1}{\pi}{\rm p.v.}\int\limits_{- \infty}^{+ \infty}
d\zeta\frac{f(\zeta)}{\zeta - \eta}\,,$$ "${\rm p.v.}$" denoting
the principal value. Equations (\ref{solup1}), (\ref{solup2})
represent the general form of the potential velocity filed
satisfying the boundary conditions at $\xi = - \infty,$ while
Eq.~(\ref{solup3}) is nothing but the Bernoulli equation. Note
that the 2D Landau-Darrieus operator $\hat{\Phi}$ is simply
expressed through the Hilbert operator $\hat{\Phi} = -
\partial\hat{H}.$ Note also that although the relation $w =
\hat{H}\tilde{u}$ between the velocity components upstream is
nonlocal, it is expressed in terms of the transverse coordinate
$\eta$ only.

\subsection{Bulk dynamics in transverse representation}\label{transvers}

Our next step is the reduction of the system of flow equations
(\ref{flow1}) -- (\ref{flow3}) to one equation in which the role of the
coordinate $\xi$ is purely parametric.
For this purpose, it is convenient to introduce the stream function
$\psi$ via
\begin{eqnarray}\label{stream1}&&
u = \frac{\partial\psi}{\partial\eta}\,, ~~w = -
\frac{\partial\psi}{\partial\xi}\,.
\end{eqnarray}
\noindent The stream function satisfies the following equation
\begin{eqnarray}\label{stream2}&&
\left(\frac{\partial}{\partial\tau} + {\bf
v}\nabla\right)\nabla^2\psi = 0\,.
\end{eqnarray}
\noindent

\subsubsection{First order approximation}

To perform the weak nonlinearity expansion, it is convenient to explicitly
extract zero-order values of the flow variables downstream
\begin{eqnarray}\label{design}&&
u = \theta + \tilde{u}\,, ~~\Pi = - \theta + 1 + \tilde{\Pi}\,.
\end{eqnarray}
\noindent
Then in the linear approximation, Eq.~(\ref{stream2}) takes the form
\begin{eqnarray}\label{stream3}&&
\left(\frac{\partial}{\partial\tau} +
\theta\frac{\partial}{\partial\xi}\right)\nabla^2\psi^{(1)} = 0\,.
\end{eqnarray}
\noindent Its general solution can be written as a superposition
of the potential and vorticity modes satisfying, respectively,
\begin{eqnarray}
\nabla^2 \psi^{(1)}_p &=& 0\,, \label{poten}\\
\left(\frac{\partial}{\partial\tau} +
\theta\frac{\partial}{\partial\xi}\right)\psi^{(1)}_v &=&
0\,.\label{vort1}
\end{eqnarray}
General solution of Eq.~(\ref{poten}) has the form analogous to
Eqs.~(\ref{solup1}), (\ref{solup2})
\begin{eqnarray}&&\label{soldown1}
u_p \equiv \theta + \tilde{u}_p = \theta + \int\limits_{-
\infty}^{+ \infty}d k~\tilde{u}_k \exp (- |k|\xi + i k\eta)\,,
\end{eqnarray}
\begin{eqnarray}&&\label{soldown2}
w_p = - \hat{H}\tilde{u}_p\,.
\end{eqnarray}
\noindent Differentiating Eq.~(\ref{vort1}) with respect to $\eta$
we obtain
\begin{eqnarray}\label{vort3}&&
\frac{1}{\theta}\frac{\partial\tilde{u}_v}{\partial\tau} =
\frac{\partial w_v}{\partial\eta}\,.
\end{eqnarray}
\noindent Next, linearizing Eq.~(\ref{flow2}) and using the above
equations for the potential and vorticity modes, one finds
\begin{eqnarray}
- \frac{1}{\theta}\frac{\partial\tilde{u}_p}{\partial\tau} +
\hat{\Phi}(\tilde{\Pi}_p + \tilde{u}_p) &=& 0\,, \label{soldown3}\\
\Pi_v &=& 0\,.\label{vort2}
\end{eqnarray}
\noindent
With the help of Eqs.~(\ref{soldown2}), (\ref{vort3}), and
(\ref{vort2}) equation (\ref{soldown3}) can be rewritten as
\begin{eqnarray}\label{vort4}&&
\hat{\Phi}\tilde{\Pi} + \frac{\partial w}{\partial\eta} -
\frac{1}{\theta}\frac{\partial\tilde{u}}{\partial\tau} = 0\,.
\end{eqnarray}
\noindent In this form, the flow equation governing dynamics
downstream contains no explicit operation with the
$\xi$-dependence of the flow variables. In other words, this
dependence is rendered purely parametric. Let us now show that
Eq.~(\ref{vort4}) can be generalized to take into account
interaction of the perturbations.

\subsubsection{Second order approximation}

At the second order, Eq.~(\ref{stream2}) takes the form
\begin{eqnarray}\label{stream5}&&
\left(\frac{\partial}{\partial\tau}
+ \theta\frac{\partial}{\partial\xi}\right)\nabla^2\psi
= - \left(\tilde{u}^{(1)}\frac{\partial}{\partial\xi}
+ w^{(1)}\frac{\partial}{\partial\eta}\right)\nabla^2\psi^{(1)}.
\end{eqnarray}
\noindent General solution of this inhomogeneous equation is the
sum of general solution of the homogeneous equation, given by
Eqs.~(\ref{soldown1})--(\ref{vort2}), and of a particular solution
$\psi_a^{(2)}$ which can be chosen to satisfy the following
equation
\begin{eqnarray}\label{stream6}&&
\left(\frac{\partial}{\partial\tau} +
\theta\frac{\partial}{\partial\xi}\right)\nabla\psi_a^{(2)}
= - {\bf v}^{(1)}\nabla^2\psi^{(1)}.
\end{eqnarray}
\noindent
Written in components, Eq.~(\ref{stream6}) has the form
\begin{eqnarray}
\frac{\partial\tilde{u}_a^{(2)}}{\partial\tau} +
\theta\frac{\partial\tilde{u}_a^{(2)}}{\partial\xi} +
w^{(1)}\left(\frac{\partial\tilde{u}^{(1)}}{\partial\eta} -
\frac{\partial w^{(1)}}{\partial\xi}\right) &=& 0\,, \\
\frac{\partial w_a^{(2)}}{\partial\tau} + \theta\frac{\partial
w_a^{(2)}}{\partial\xi} -
\tilde{u}^{(1)}\left(\frac{\partial\tilde{u}^{(1)}}{\partial\eta}
- \frac{\partial w^{(1)}}{\partial\xi}\right) &=& 0\,.
\label{stream71}
\end{eqnarray}
\noindent Next, retaining the second order terms in the Euler
equation (\ref{flow2}) and using Eqs.~(\ref{poten}),
(\ref{vort3}), (\ref{stream71}) one can obtain the following
relation between the velocity and pressure fields
\begin{eqnarray}\label{stream72}&&
\frac{\partial w_p}{\partial\tau} + \theta\frac{\partial
w_p}{\partial\xi} +
\frac{\partial}{\partial\eta}\left(\theta\tilde{\Pi} +
\frac{\tilde{u}^2 + w^2}{2}\right) = 0\,.
\end{eqnarray}
\noindent Finally, taking into account explicit structure of the
potential mode and Eqs.~(\ref{soldown2}), (\ref{vort3}),
(\ref{stream71}) it is not difficult to verify that
Eq.~(\ref{stream72}) can be rewritten in terms of the sum
$(w,\tilde{u}) = {\bf v}_p^{(1)} + {\bf v}_v^{(1)} + {\bf
v}_a^{(2)}$ as follows
\begin{eqnarray}\label{stream8}&&
\frac{\partial\tilde{u}}{\partial\tau} - \theta\frac{\partial
w}{\partial\eta} - \hat{\Phi}\left(\theta\tilde{\Pi} +
\frac{\tilde{u}^2 + w^2}{2}\right) +
w\left(\frac{\partial\tilde{u}}{\partial\eta} +
\frac{1}{\theta}\frac{\partial w}{\partial\tau} +
\frac{\partial\tilde{\Pi}}{\partial\eta}\right) = 0\,,
\end{eqnarray}
\noindent
which is the transverse representation of the flow equations at the
second order of nonlinearity we are looking for.

\section{Nonlinear equation for the flame front}\label{nnequation}

Now we can show that the set of equations (\ref{conserv1}) --
(\ref{evolution}), (\ref{solup2}), (\ref{solup3}), and
(\ref{stream8}) can be reduced to one equation for the function
$f(\eta,\tau).$ To this end, it remains only to rewrite the right
hand sides of the conservation laws and evolution equation in the
form in which transition on the flame surface ($\xi\to
f(\eta,\tau)$) is performed {\it after} all differentiations,
again bringing the latter to the transverse form. This is easily
done using the continuity equation (\ref{flow1}) and taking into
account potentiality of the flow upstream
\begin{eqnarray}
\frac{d w_{-}}{d\eta} &=& \left(\frac{\partial
w}{\partial\eta}\right)_{-} + \left(\frac{\partial w
}{\partial\xi}\right)_{-}\cdot f' = \left(\frac{\partial
w}{\partial\eta}\right)_{-} + \left(\frac{\partial u
}{\partial\eta}\right)_{-}\cdot f'\,, \label{auxil1}\\
 \hat{D}w_{-} &=& \left(\hat{D}w\right)_{-} +
\left(\frac{\partial w }{\partial\xi}\right)_{-}\cdot\hat{D}f =
\left(\hat{D}w\right)_{-} + \left(\frac{\partial u
}{\partial\eta}\right)_{-}\cdot\hat{D}f\,, \label{auxil2}\\
\hat{D}u_{-} &=& \left(\hat{D}u\right)_{-} + \left(\frac{\partial
u }{\partial\xi}\right)_{-}\cdot\hat{D}f =
\left(\hat{D}u\right)_{-} - \left(\frac{\partial w
}{\partial\eta}\right)_{-}\cdot\hat{D}f\,. \label{auxil3}
\end{eqnarray}

After having done this, one sees that the knowledge of explicit
$\xi$-dependence of the flow variables turns out to be
unnecessary. Roughly speaking, the $\xi$-dependence of a function
$F(\xi,\eta,\tau)$ describing the flame front shape is known in
advance, since the equation $F(\xi,\eta,\tau) = 0$ can always be
brought into the form $\xi - f(\eta,\tau) = 0$ (with $f$
many-valued, in general). To determine the flame front evolution,
therefore, it is sufficient to find only $(\eta,\tau)$-dependence
of the functions involved. To put this intuitive reasoning in the
formal way, it is convenient to introduce separate designations
for the up- and downstream velocity and pressure fields,
distinguishing them by the {\it super}scripts $"-"$ and $"+"$,
respectively. Then, setting $\xi = f(\eta,\tau),$
equations~(\ref{solup2}), (\ref{solup3}), (\ref{stream8}) together
with the conservation laws (\ref{conserv1}) -- (\ref{conserv3})
and evolution equation (\ref{evolution}) can be rewritten
identically as follows
$$\left(
\begin{array}{rcl}
w^{-}&=&\hat{H}\tilde{u}^{-}\\
\frac{\partial \tilde{u}^{-}}{\partial\tau} + \hat{\Phi}\Pi^{-}
+ \frac{\hat{\Phi}}{2}\{(u^{-})^2 + (w^{-})^2\} &=& 0\\
\frac{\partial\tilde{u}^{+}}{\partial\tau}
- \theta\frac{\partial w^{+}}{\partial\eta}
- \hat{\Phi}\left(\theta\tilde{\Pi}^{+}
+ \frac{(\tilde{u}^{+})^2 + (w^{+})^2}{2}\right)
&=& - w^{+}\left(\frac{\partial\tilde{u}^{+}}{\partial\eta}
+ \frac{1}{\theta}\frac{\partial w^{+}}{\partial\tau}
+ \frac{\partial\tilde{\Pi}^{+}}{\partial\eta}\right) \\
\tilde{u}^{+} - \tilde{u}^{-} - \frac{\partial f}{\partial\eta}
(w^{+} - w^{-})&=&\frac{\theta - 1}{2}
\left(\frac{\partial f}{\partial\eta}\right)^2\\
w^{+} - w^{-} + \frac{\partial f}{\partial\eta}(u^{+} - u^{-})&=&
\varepsilon\ln\theta\left(\frac{\partial w^{-}}{\partial\tau}
+ \frac{\partial^2 f}{\partial\eta\partial\tau}\right)\\
\tilde{\Pi}^{+} - \Pi^{-}&=& \varepsilon (\theta - 1)
\frac{\partial^2 f}{\partial\eta^2}
+ \varepsilon\ln\theta\frac{\partial^2 f}{\partial\tau^2}\\
\tilde{u}^{-} - \frac{\partial f}{\partial\eta} w^{-}
- \frac{\partial f}{\partial\tau}&=&
\frac{1}{2}\left(\frac{\partial f}{\partial\eta}\right)^2
- \varepsilon\frac{\theta\ln\theta}{\theta - 1}
\left(\frac{\partial\tilde{w}^{-}}{\partial\eta}
+ \frac{\partial^2 f}{\partial\eta^2}\right)\\
\end{array}
\right)_{\xi = f(\eta,\tau)}\eqno (*)$$

Suppose we found a solution $f = f(\eta,\tau),$ ${\bf v}^{-} =
{\bf v}^{-}(\xi,\eta,\tau),$ ${\bf v}^{+} = {\bf
v}^{+}(\xi,\eta,\tau),$ etc. of the set of equations in the large
brackets in ($*$). Then, in particular, these equations are
satisfied for $\xi = f(\eta,\tau).$ On the other hand, since no
operation involving $\xi$ appears in these equations, the function
$f(\eta,\tau)$ is independent of a particular form of
$\xi$-dependence of the flow variables. For the purpose of
deriving an equation for $f(\eta,\tau),$ it is most convenient to
work with the particular solution in which all the functions are
$\xi$-independent, and to omit the large brackets in ($*$).
Therefore, we can replace the above set of equations by the
following
\begin{eqnarray}
\omega^{-}&=&\hat{H}\tilde{\upsilon}^{-}\label{eq1}\\
\frac{\partial \tilde{\upsilon}^{-}}{\partial\tau} + \hat{\Phi}\pi^{-}
+ \frac{\hat{\Phi}}{2}\{(\upsilon^{-})^2 + (\omega^{-})^2\} &=& 0\label{eq2}\\
\frac{\partial\tilde{\upsilon}^{+}}{\partial\tau}
- \theta\frac{\partial \omega^{+}}{\partial\eta}
- \hat{\Phi}\left(\theta\tilde{\pi}^{+}
+ \frac{(\tilde{\upsilon}^{+})^2 + (\omega^{+})^2}{2}\right)
&=& - \omega^{+}\left(\frac{\partial\tilde{\upsilon}^{+}}{\partial\eta}
+ \frac{1}{\theta}\frac{\partial \omega^{+}}{\partial\tau}
+ \frac{\partial\tilde{\pi}^{+}}{\partial\eta}\right) \label{eq3}\\
\tilde{\upsilon}^{+} - \tilde{\upsilon}^{-} - \frac{\partial f}{\partial\eta}
(\omega^{+} - \omega^{-})&=&\frac{\theta - 1}{2}
\left(\frac{\partial f}{\partial\eta}\right)^2\label{eq4}\\
\omega^{+} - \omega^{-} + \frac{\partial f}{\partial\eta}
(\upsilon^{+} - \upsilon^{-})&=&
\varepsilon\ln\theta\left(\frac{\partial \omega^{-}}{\partial\tau}
+ \frac{\partial^2 f}{\partial\eta\partial\tau}\right)\label{eq5}\\
\tilde{\pi}^{+} - \pi^{-}&=& \varepsilon (\theta - 1)
\frac{\partial^2 f}{\partial\eta^2}
+ \varepsilon\ln\theta\frac{\partial^2 f}{\partial\tau^2}\label{eq6}\\
\tilde{\upsilon}^{-} - \frac{\partial f}{\partial\eta} \omega^{-}
- \frac{\partial f}{\partial\tau}&=&
\frac{1}{2}\left(\frac{\partial f}{\partial\eta}\right)^2
- \varepsilon\frac{\theta\ln\theta}{\theta - 1}
\left(\frac{\partial\omega^{-}}{\partial\eta}
+ \frac{\partial^2 f}{\partial\eta^2}\right)\label{eq7},
\end{eqnarray}
where $\upsilon, \omega,$ and $\pi$ are the
$\xi$-independent counterparts of the flow variables $u, w,$ and
$\Pi,$ respectively, and
\begin{eqnarray}\label{decompose}
\upsilon^{-} \equiv 1 + \tilde{\upsilon}^{-},
~~\upsilon^{+}
\equiv \theta + \tilde{\upsilon}^{+}.
\end{eqnarray}
\noindent The fact that now the function $f(\eta,\tau)$ does not enter
the arguments of these variables allows us to avoid expanding them
in powers of $f,$ employed in Ref.~\cite{zhdanov}.
In fact, such an expansion is irrelevant to the issue, since all the
equations governing flame propagation are invariant with respect to the
space translations, and therefore, all terms containing powers of
undifferentiated $f$ should appear in invariant combinations in the final
equation for $f.$ In view of this invariance, the function $f$ itself
does not need to be small even if the front is only weakly curved.
We thus see that the $f$-dependence of the flow variables through their
arguments must eventually cancel out in some way in any case.

Now, the system of Eqs.~(\ref{eq1}) -- (\ref{eq7}) can be
transformed into one equation for the function $f(\eta,\tau).$ To
simplify the derivation, the stabilizing effects due to the finite
flame thickness will be taken into account only in the linear
approximation. As we already mentioned in the Introduction, these
effects are mainly taken into account in order to provide the
short wavelength cutoff for the spectrum of the flame front
perturbations. In view of this, the nonlinear corrections in the
$\varepsilon$-terms are of little interest. Using Eq.~(\ref{eq1}),
the ``evolution equation'' (\ref{eq7}) can be rewritten within the
accuracy of the second order
\begin{eqnarray}\label{eq71}
\tilde{\upsilon}^{-} = \frac{\partial f}{\partial\tau} +
\frac{1}{2}\left(\frac{\partial f}{\partial\eta}\right)^2
+ \frac{\partial f}{\partial\eta}\hat{H}\frac{\partial f}{\partial\tau}
- \varepsilon\frac{\theta\ln\theta}{\theta - 1}
\left(\hat{H}\frac{\partial^2 f}{\partial\eta\partial\tau}
+ \frac{\partial^2 f}{\partial\eta^2}\right).
\end{eqnarray}
\noindent
Next, solving Eqs.~(\ref{eq4}) -- (\ref{eq6}) with respect to $\upsilon^{+},$
$\omega^{+},$ $\pi^{+},$ and substituting the results into Eq.~(\ref{eq3}),
one obtains an equation for the upstream variables $\tilde{\upsilon}^{-},$
$\omega^{-},$ $\pi^{-},$ which can be further reduced to an equation for
$\tilde{\upsilon}^{-}$ alone using Eqs.~(\ref{eq1}), (\ref{eq2}).
Upon substituting the expression (\ref{eq71}) into the latter equation,
one arrives at the nonlinear equation for the function $f(\eta,\tau),$
which we do not write explicitly because of its great complexity.
It can be highly simplified using the first order LD-equation
\begin{eqnarray}\label{ld}
\frac{\partial f}{\partial\tau} = \nu\hat{\Phi}f,
~~\nu = \frac{\theta}{\theta + 1}
\left(\sqrt{1 + \theta - \frac{1}{\theta}} - 1\right),
\end{eqnarray}
\noindent in the second order terms and in the terms containing
$\varepsilon.$ The above value for $\nu$ corresponds to the
exponentially growing solution in the LD-theory. The nonlinear
equation thus takes the form
\begin{eqnarray}\label{main1}
\frac{\theta + 1}{2\theta}\hat{\Phi}^{-1}\frac{\partial^2
f}{\partial\tau^2} + \frac{\partial f}{\partial\tau} -
\frac{\theta - 1}{2}\hat{\Phi}f +
\tilde{\alpha}\hat{\Phi}^{-1}\frac{\partial}{\partial\tau}
\left(\frac{\partial f}{\partial\eta}\right)^2 +
\tilde{\beta}\left(\frac{\partial f}{\partial\eta}\right)^2 +
\tilde{\gamma}\left(\hat{\Phi}f\right)^2 +
\varepsilon\tilde{\delta}\frac{\partial^2 f}{\partial\eta^2} =
0\,,
\end{eqnarray}
\noindent
where
\begin{eqnarray}\label{main1coef}&&
\tilde{\alpha} = \frac{(\nu + 1)^2}{4\theta^2} + \frac{3\nu +
1}{4} - \frac{\nu (\nu + 1)}{4\theta}\,, \nonumber\\&&
\tilde{\beta} = \nu + \frac{1}{2} + \frac{\theta -
1}{4\theta}\left[(\nu + 1)^2 - \theta \right]\,, \nonumber\\&&
\tilde{\gamma} = \frac{\theta - 1}{4\theta}\nu^2\,, \nonumber\\&&
\tilde{\delta} = - \frac{(\nu + 1)\ln\theta}{\theta - 1}
\left(\frac{\theta + 1}{2}\nu + \theta\right) -
\frac{\nu}{2}\ln\theta - \frac{\theta - 1}{2}\,. \nonumber
\end{eqnarray}
\noindent
Following Zhdanov and Trubnikov \cite{zhdanov}, Eq.~(\ref{main1})
can be further simplified by rewriting its linear part in the form
\begin{eqnarray}\label{aux}
\frac{\theta + 1}{2\theta}\hat{\Phi}^{-1}\frac{\partial^2
f}{\partial\tau^2} + \frac{\partial f}{\partial\tau} -
\frac{\theta - 1}{2}\hat{\Phi}f = \left(\frac{\theta +
1}{2\theta}\hat{\Phi}^{-1} \frac{\partial}{\partial\tau} +
\frac{\theta - 1}{2\nu}\right) \left(\frac{\partial}{\partial\tau}
- \nu\hat{\Phi}\right) f\,.
\end{eqnarray}
\noindent
Since we are only interested in the development of unstable modes of the
front perturbations, which satisfy Eq.~(\ref{ld}), we can transform
the first factor in the right hand side of Eq.~(\ref{aux}) as follows
\begin{eqnarray}
\frac{\theta +
1}{2\theta}\hat{\Phi}^{-1}\frac{\partial}{\partial\tau} +
\frac{\theta - 1}{2\nu} \to \frac{\theta + 1}{2\theta}\nu +
\frac{\theta - 1}{2\nu} = \sqrt{1 + \theta - \frac{1}{\theta}}\ .
\nonumber
\end{eqnarray}
\noindent

Therefore, Eq.~(\ref{main1}) becomes
\begin{eqnarray}\label{main2}
 \frac{\partial f}{\partial\tau} - \nu\hat{\Phi}f
+ \alpha\hat{\Phi}^{-1}\frac{\partial}{\partial\tau}
\left(\frac{\partial f}{\partial\eta}\right)^2 +
\beta\left(\frac{\partial f}{\partial\eta}\right)^2 +
\gamma\left(\hat{\Phi}f\right)^2 +
\varepsilon\delta\frac{\partial^2 f}{\partial\eta^2} = 0\,,
\end{eqnarray}
\noindent where $$(\alpha, \beta, \gamma, \delta) =
\frac{(\tilde{\alpha}, \tilde{\beta}, \tilde{\gamma},
\tilde{\delta})}{\sqrt{1 + \theta - \frac{1}{\theta}}}\,.$$

Finally, we would like to comment on the range of validity of the
derived equation. Generally speaking, Eq.~(\ref{main2}) is only
applicable for description of the early stages of development of
the LD-instability, since it is obtained under the assumption of
smallness of the front slope. Even if the flame evolution is such
that it smoothly ends up with the formation of a stationary
configuration (instead of spontaneous turbulization), this
assumption becomes generally invalid whenever the process of flame
propagation is close to the stationary regime. In fact, it can be
easily shown that the assumptions of stationarity and weak
nonlinearity contradict each other (detailed discussion of this
point can be found in Ref.~\cite{kazakov}). Incidentally, the fact
that the transition to the stationary regime in Eq.~(\ref{main2})
is formally incorrect is clearly seen from its derivation given
above. Namely, the stationary form of this equation depends on the
way the first order relation (\ref{ld}) is used in the second
order terms before time derivatives are omitted. Only in the case
of small gas expansion ($\theta\to 1$) is weak nonlinearity
approximation justified at all stages of development of the
LD-instability, in which case Eq.~(\ref{main2}) goes over to the
well-known Sivashinsky equation \cite{siv2}
\begin{eqnarray}\label{siv}
\frac{\partial f}{\partial\tau} + \frac{1}{2}\left(\frac{\partial
f}{\partial\eta}\right)^2 = \frac{\theta - 1}{2}\hat{\Phi}f -
\varepsilon\frac{\partial^2 f}{\partial\eta^2}\,,
\end{eqnarray}
\noindent since $$\beta = \frac{1}{2} + O((\theta - 1 )^2)\,,
~~\gamma = O((\theta - 1)^3)\,,~~\delta = - 1 + O(\theta - 1)\,,
~~\nu = \frac{\theta - 1}{2} + O((\theta - 1)^2)\,.$$ In this
respect, a natural question arises as to what extent equation
(\ref{main2}) is actually valid when $\theta$ is arbitrary. Since
the structure of higher order terms of the power expansion is
unknown, it is very difficult to give even a rough estimate.
Leaving this question aside, we will simply {\it assume} in what
follows, that this equation is formally valid for all times. It
will be shown below that at least in the case of flame propagation
in narrow tubes, solutions to the stationary version of
Eq.~(\ref{main2}) are in reasonable agreement with the results of
numerical experiments \cite{golberg} for flames with the gas
expansion coefficient up to $\theta \sim 3$.

\section{The pole decomposition}\label{decomposition}

As in the case of ZT-equation, development of the LD-instability
of a plane flame can be described in terms of the pole dynamics.
To show this, we first perform the following nonlinear
transformation
\begin{eqnarray}\label{change}
f\to\phi = f + \alpha \hat{\Phi}^{-1} \left(\frac{\partial
f}{\partial\eta}\right)^2\,.
\end{eqnarray}
\noindent
In terms of the new function $\phi,$ Eq.~(\ref{main2}) takes the form,
within the accuracy of the second order,
\begin{eqnarray}\label{main3}
\frac{\partial \phi}{\partial\tau} - \nu\hat{\Phi}\phi + (\beta +
\nu\alpha)\left(\frac{\partial \phi}{\partial\eta}\right)^2 +
\gamma\left(\hat{\Phi}\phi\right)^2 +
\varepsilon\delta\frac{\partial^2 \phi}{\partial\eta^2} = 0\,.
\end{eqnarray}
\noindent

Spatially periodic (with period $2 b$) solutions of equations of
the type (\ref{main3}) can be found using the following pole
decomposition \cite{joulin,henon}
\begin{eqnarray}\label{ansatz}
\phi(\eta,\tau) = \phi_0(\tau) + a \sum\limits_{k=1}^{2 P}
\ln\sin\left[\frac{\pi}{2 b}(\eta - \eta_k(\tau))\right]\,,
\end{eqnarray}
\noindent where the value of the amplitude $a$ as well as dynamics of
the complex poles $\eta_k(\tau),$ $k = 1,...,2P$ are to be determined
substituting this anzats into Eq.~(\ref{main3}).
Since the function $\phi(\eta,\tau)$ is real for
real $\eta,$ the poles come in conjugate pairs; $P$ is the number
of the pole pairs. Requiring the $2 b$-periodic solutions to be
symmetric with respect to the reflection $\eta \to - \eta,$ one
can obtain periodic as well as non-periodic solutions to
Eq.~(\ref{main3}) in the domain $\eta \in (0,~b)$, satisfying
the conditions $\frac{\partial\phi}{\partial\eta}(0,\tau)
= \frac{\partial\phi}{\partial\eta}(b,\tau) = 0,$ describing
flame propagation in a tube of width $b$ with ideal walls.

Using the formulae \cite{joulin}
\begin{eqnarray}&&\label{fjoul}
\hat{H}\frac{\partial\phi}{\partial\eta} = - \frac{\pi a}{2
b}\sum_{k = 1}^{2 P}\left\{1 + i~{\rm sign}({\rm
Im}~\eta_k)\cot\left[\frac{\pi}{2 b}(\eta -
\eta_k)\right]\right\}\,, ~{\rm sign}(x) \equiv
\frac{x}{|x|}\,,\nonumber\\&& \cot x \cot y = -1 + \cot(x - y
)(\cot y - \cot x)\,,
\end{eqnarray}
\noindent it is not difficult to verify that Eq.~(\ref{main3})
is satisfied by $\phi(\eta,\tau)$ taken in the form of Eq.~(\ref{ansatz}),
provided that
\begin{eqnarray}&&\label{solution1}
a = \varepsilon\delta\chi, \nonumber\\&&
\frac{\partial\phi_0}{\partial\tau} = \chi(\sigma^2 P^2 -
\nu\sigma P)\,, \nonumber
\end{eqnarray}
\noindent and the poles $\eta_k(\tau),$ $k = 1,...,2P,$ satisfy
\begin{eqnarray}&&\label{solution2}
\frac{\partial\eta_k}{\partial\tau} + i~{\rm sign}({\rm
Im}~\eta_k) \left(\nu + 2 P\gamma\sigma\chi\right) \nonumber\\&& -
\sigma\chi\sum\limits_{m = 1 \atop m\ne k}^{2 P} \{\gamma{\rm
sign}({\rm Im}~\eta_k~{\rm Im}~\eta_m) - \beta - \nu\alpha\}
\cot\left[\frac{\pi}{2 b}(\eta_k - \eta_m)\right] = 0\,,
\end{eqnarray}
\noindent where the following notation is introduced
$$\sigma \equiv - \frac{\varepsilon\delta\pi}{b}>0\,,
~~\chi\equiv (\beta + \nu\alpha - \gamma)^{-1}\ .$$

Since the application of pole decomposition to Eq.~(\ref{main3})
is quite similar to that given in Refs.~\cite{joulin,henon}, we
will present below only final results, referring the reader to
these works for more detail.

Following Ref.~\cite{henon}, we first consider two poles ($\eta_1,
\eta_2$) in the same half plane of the complex $\eta$, which are
fairly close to each other, so that their dynamics is unaffected
by the rest. Then one has from Eq.~(\ref{solution2})
$$\frac{\partial}{\partial\tau}(\eta_1 - \eta_2) =
\frac{4\varepsilon\delta}{\eta_1 - \eta_2}\,,$$ which indicates
that the poles attract each other in the horizontal direction
(parallel to the real axis), and repel in the vertical direction,
tending to form alignments parallel to the imaginary axis.
Furthermore, assuming that the pole dynamics ends up with the
formation of such a ``coalescent'' stationary configuration, and
using the fact that $\gamma<\beta + \nu\alpha$ (it is not
difficult to verify that actually $\gamma/(\beta +
\nu\alpha)<1/3$), the following upper bound on the number of pole
pairs $P$ can be easily obtained from Eq.~(\ref{solution2}) $$P
\le \frac{1}{2} + \frac{\nu}{2\sigma}\ .$$ Still, for sufficiently
wide tubes (such that $\sigma<\nu/3$), the solution (\ref{ansatz})
is not unique: different solutions corresponding to different
numbers $P$ of poles are possible. To find the physical ones, it
is necessary to perform the stability analysis. Noting that the
functional structure of Eq.~(\ref{main3}) is very similar to that
of Eq.~(\ref{siv}), the stability analysis of
Refs.~\cite{siv3}--\cite{siv5}, where it was carried out for the
Sivashinsky equation, will be carried over the present case.
According to this analysis, for a given not-too-wide tube, there
is only one (neutrally) stable solution. This solution corresponds
to the number of poles that provides maximal flame velocity, i.e.,
$$P_{{\rm m}} = {\rm Int} \left(\frac{\nu}{2\sigma} + \frac{1}{2}\right),$$
${\rm Int}(x)$ denoting the integer part of $x.$ Thus, the flame
velocity increase $W_s\equiv \left|\partial\phi_0/\partial\tau
\right|$ of the stable solution can be written as
\begin{eqnarray}&&\label{solution3}
W_s = 4 W_{{\rm m}}\frac{\sigma P_{{\rm m}}}{\nu}\left(1 -
\frac{\sigma P_{{\rm m}}}{\nu} \right),
\end{eqnarray}
\noindent where
\begin{eqnarray}\label{wmax}
W_{{\rm m}} = \frac{\chi\nu^2}{4}\,.
\end{eqnarray}

It is seen from Eq.~(\ref{main3}) that the spectrum of front
perturbations is effectively cut off at the wavelength
\begin{eqnarray}\label{cutoff}
\lambda = \frac{2\pi\varepsilon|\delta|}{\nu}\,,
\end{eqnarray}
representing the characteristic dimension of the flame cellular
structure.

\begin{figure}
\epsfxsize=8,5cm\epsfbox{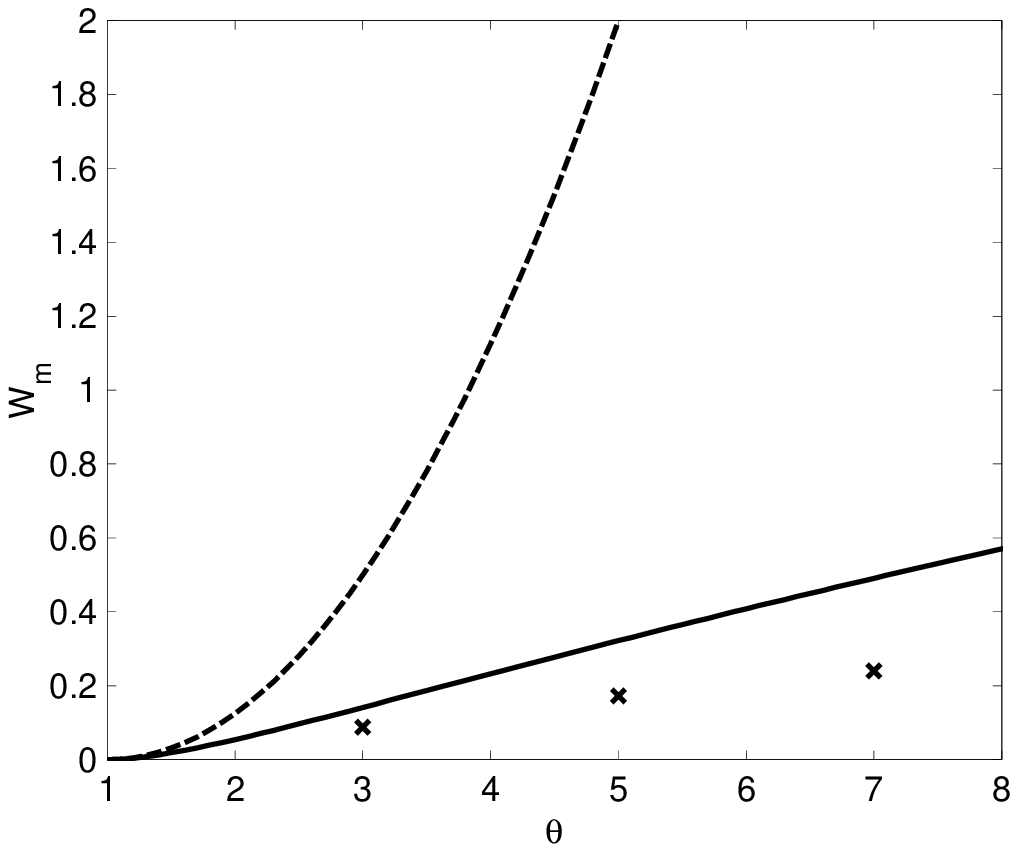} \caption{Maximal flame
velocity increase $W_{{\rm m}}$ versus the gas expansion
coefficient, given by Eq.~(\ref{wmax}); the marks are according to
Ref.~[8]. Accuracy of the experimental results is about
20\%.}\label{fig1}
\end{figure}

\begin{figure}
\epsfxsize=8,5cm\epsfbox{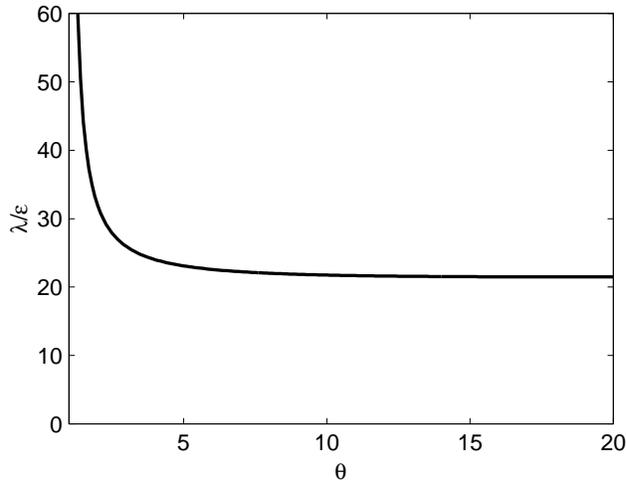} \caption{The cut-off
wavelength scaled on the flame thickness versus the expansion
coefficient, given by Eq.~(\ref{cutoff}).}\label{fig2}
\end{figure}

Fig.~\ref{fig1} compares the theoretical dependence of the maximal
flame velocity increase $W_{{\rm m}}$ on the gas expansion
coefficient, given by Eq.~(\ref{wmax}), with the results of
numerical experiments \cite{golberg}. For comparison, we show also
the corresponding dependence calculated on the basis of the
Sivashinsky equation. Dependence of the effective cut-off
wavelength on the expansion coefficient is shown in
Fig.~\ref{fig2}. We see that even beyond of its range of
applicability, Eq.~(\ref{main3}) provides reasonable qualitative
description of flames with the expansion coefficients of practical
interest. Complete investigation of the LD-instability on the
basis of this equation will be given elsewhere.

\section{Conclusions}\label{conclud}

The main result of our work is the nonlinear non-stationary
equation (\ref{main1}) which describes development of the
Landau-Darrieus instability in the second order of nonlinearity.
We have derived this equation on the basis of the only assumption
of smallness of the flame front slope. Thus, nonlinear evolution
of the front perturbations generally obeys Eq.~(\ref{main1}) which
takes even simpler form (\ref{main3}) if one is only interested in
dynamics of the exponentially growing LD-solutions. It is
important to stress that no assumption concerning the value of the
flame front position has been used in the derivation. Therefore,
Eq.~(\ref{main1}) can be applied not only to plane flames, but
also to the problem of unstable evolution of any flame
configuration, provided that the front slope is small.

It is also worth of emphasis that Eq.~(\ref{main1}) is obtained
without any assumptions about the structure of the gas flow
downstream. Thus, this equation is the direct consequence of the
exact hydrodynamic equations for the flow fields in the bulk, and
conservation laws at the flame front. We would like to stress also
once again that the universal form of Eq.~(\ref{main1}) is the
distinguishing property of the second order approximation. In the
general case, equation for the flame front should contain also
information about the flow of the burnt matter in the bulk.
Indeed, as was shown in Ref.~\cite{kazakov}, the boundary
conditions for the burnt matter are invoked in the course of
derivation of the equation already at the third order. This
universality of Eq.~(\ref{main1}) allows it to be widely applied
to the study of flames with arbitrary front configuration,
propagating in tubes with complex geometries.

\begin{acknowledgements}
We are grateful to V.~Lvov and G.~Sivashinsky for interesting
discussions. This research was supported in part by Swedish
Ministry of Industry (Energimyndigheten, contract P 12503-1), by
the Swedish Research Council (contract E5106-1494/2001), and by
the Swedish Royal Academy of Sciences. Support form the STINT
Fellowship program is also gratefully acknowledged.
\end{acknowledgements}

\end{document}